\documentclass[]{interact}
\usepackage{adjustbox}
\usepackage[natbibapa,nodoi]{apacite}
\usepackage{color,soul}

\usepackage[hidelinks]{hyperref}  
\usepackage{amsmath}
\usepackage[utf8x]{inputenc}
\usepackage{graphicx}
\usepackage{subcaption}
\usepackage{epstopdf}

\hypersetup{
    colorlinks=false,
    linkcolor=blue,
    filecolor=magenta,      
    urlcolor=cyan,
    pdfpagemode=FullScreen
    }

\usepackage[longnamesfirst,sort]{natbib}
\bibpunct[, ]{(}{)}{;}{a}{,}{,}
\renewcommand\bibfont{\fontsize{10}{12}\selectfont}

\theoremstyle{plain}

\theoremstyle{definition}

\usepackage{booktabs}
\usepackage{multirow}
\theoremstyle{remark}

\usepackage{lineno}

\usepackage{threeparttable} 

\newcommand{\ie}{i.\,e.,\ }

\begin{document}




\title{Enhancing Hybrid Eye Typing Interfaces with Word and Letter Prediction: A Comprehensive Evaluation}


\author{
\name{Zhe Zeng\textsuperscript{a}\thanks{CONTACT Zhe Zeng. Email: zhe.zeng@mms.tu-berlin.de}, 
Xiao Wang\textsuperscript{a},
Felix Wilhelm Siebert\textsuperscript{b},
and Hailong Liu\textsuperscript{c}}
\affil{
\textsuperscript{a}Human-Machine Systems, Technical University of Berlin, Germany;\\
\textsuperscript{b}Department of Technology, Management, and Economics, Technical University of Denmark, Denmark;\\
\textsuperscript{c}Graduate School of Science and Technology, Nara Institute of Science and Technology, Japan}}

\maketitle

\begin{abstract}
Eye typing interfaces enable a person to enter text into an interface using only their own eyes. But despite the inherent advantages of touchless operation and intuitive design, such eye-typing interfaces often suffer from slow typing speeds, resulting in slow words per minute (WPM) counts. 
In this study, we add word and letter prediction to the eye-typing interface and investigate users' typing performance as well as their subjective experience while using the interface. 
In experiment 1, we compared three typing interfaces with letter prediction (LP), letter+word prediction (L+WP), and no prediction (NoP), respectively. 
We found that the interface with L+WP achieved the highest average text entry speed (5.48 WPM), followed by the interface with LP (3.42 WPM), and the interface with NoP (3.39 WPM). 
Participants were able to quickly understand the procedural design for word prediction and perceived this function as very helpful. 
Compared to LP and NoP, participants needed more time to familiarize themselves with L+WP in order to reach a plateau regarding text entry speed.
Experiment 2 explored training effects in L+WP interfaces. Two moving speeds were implemented: slow (6.4$^\circ$/s same speed as in experiment~1) and fast (10$^\circ$/s). The study employed a mixed experimental design, incorporating moving speeds as a between-subjects factor, to evaluate its influence on typing performance throughout 10 consecutive training sessions.
The results showed that the typing speed reached 6.17 WPM for the slow group and 7.35 WPM for the fast group after practice. Overall, the two experiments show that adding letter and word prediction to eye-typing interfaces increases typing speeds. We also find that more extended training is required to achieve these high typing speeds.

\end{abstract}


\section{Introduction}
Eye-tracking equipment facilitates the registration of eye position and eye movement data. 
The technology has been used in human-computer interaction, where the eye gaze can serve as an invisible cursor and indicate the current position of the user's attention on a graphical user interface. 
Using eye-gaze as an input, applications such as PIN entry~\citep{10.1145/2857491.2857527}, wheelchair control \citep{7394111}, eye typing \citep{majaranta2002twenty}, assistance in browsing artwork imagery~\citep{dondi2022gaze}, and gaze vending interface \citep{zeng2023GaVe} have been developed. 
Gaze input offers several advantages over traditional touch-based input. As one example, gaze input supports people with disabilities in using devices when other input paradigms are not accessible to them.
Additionally, as a touchless interaction modality, gaze input reduces direct physical contact and creates a more sanitary environment. 
It also adds a degree of privacy, as gaze input is harder to observe than touch input, and hence can protect users from shoulder surfing. 
In recent years, the gaze position as the main variable for gaze-interaction systems can be estimated using off-the-shelf RGB cameras~\citep{zhang2019evaluation}. This has considerably reduced the cost of gaze-based interfaces, as simple webcams can be used for gaze input registration. Since webcams are already integrated into a large number of devices, the possibilities for using gaze-based applications have increased. 
Last but not least, eye gaze input is also showing increasing potential in head-mounted devices.

While eye-assisted text input has been extensively studied over the years, the task of gaze-only text entry remains a persistent challenge.
First, the requirement for complex calibration of the eye-tracking system to the individual user is a frequently raised challenge. 
Most QWERTY-based eye typing systems require calibration before use, which hinders new users' spontaneous interactions with the system. 
And even if a device is calibrated for an individual user, a re-calibration is often required after using the system for a certain period of time~\citep{feitchi2017}. 
In addition to the accuracy/calibration challenge of eye-tracking devices, noisy data from miniature eye movements reduce the accuracy of the gaze position estimation and restrict the interface design. In other words, it is not easy to precisely select an actionable object on an interface when the accuracy is low, hence, the size of all objects needs to be increased. This in turn can limit the number of actionable objects on an interface.

\begin{figure*}
    \centering
        \includegraphics[width=0.49\linewidth]{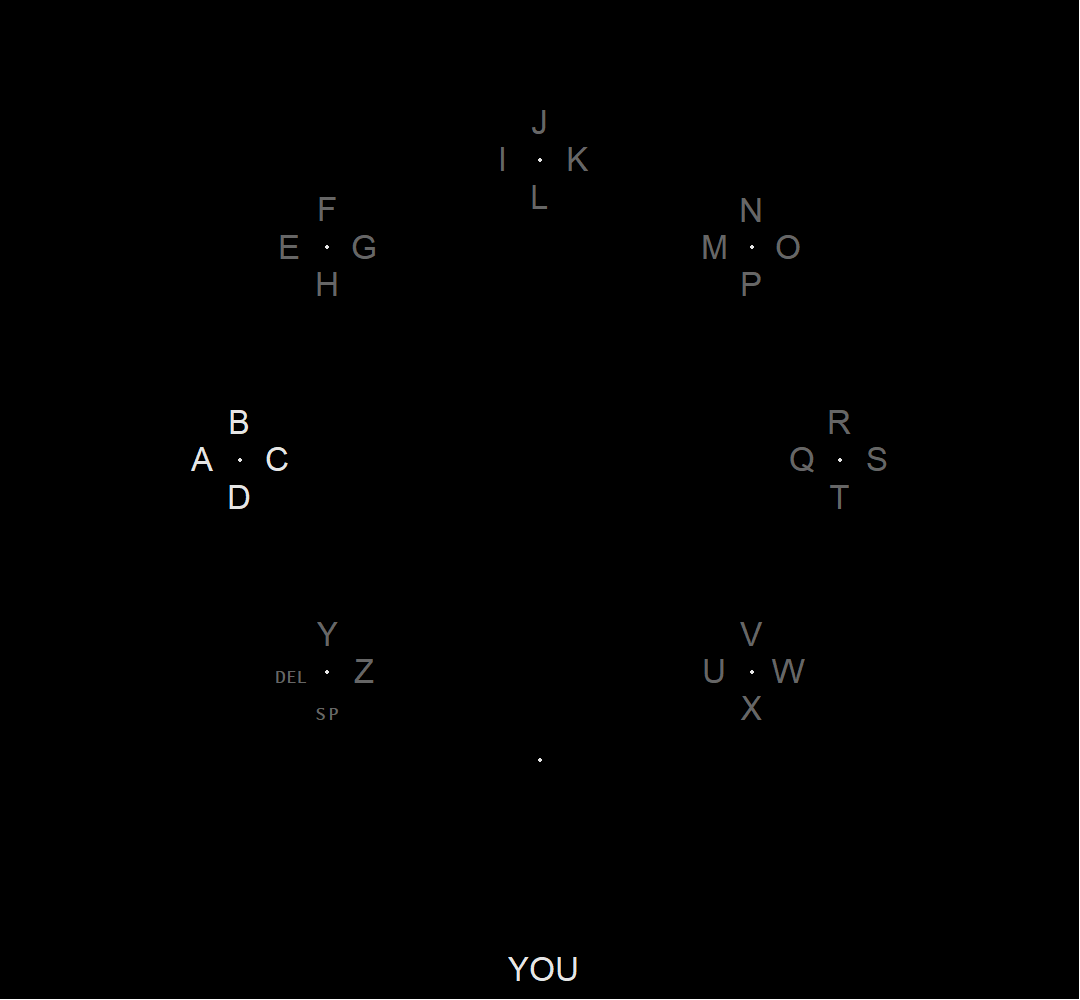}
        \includegraphics[width=0.49\linewidth]{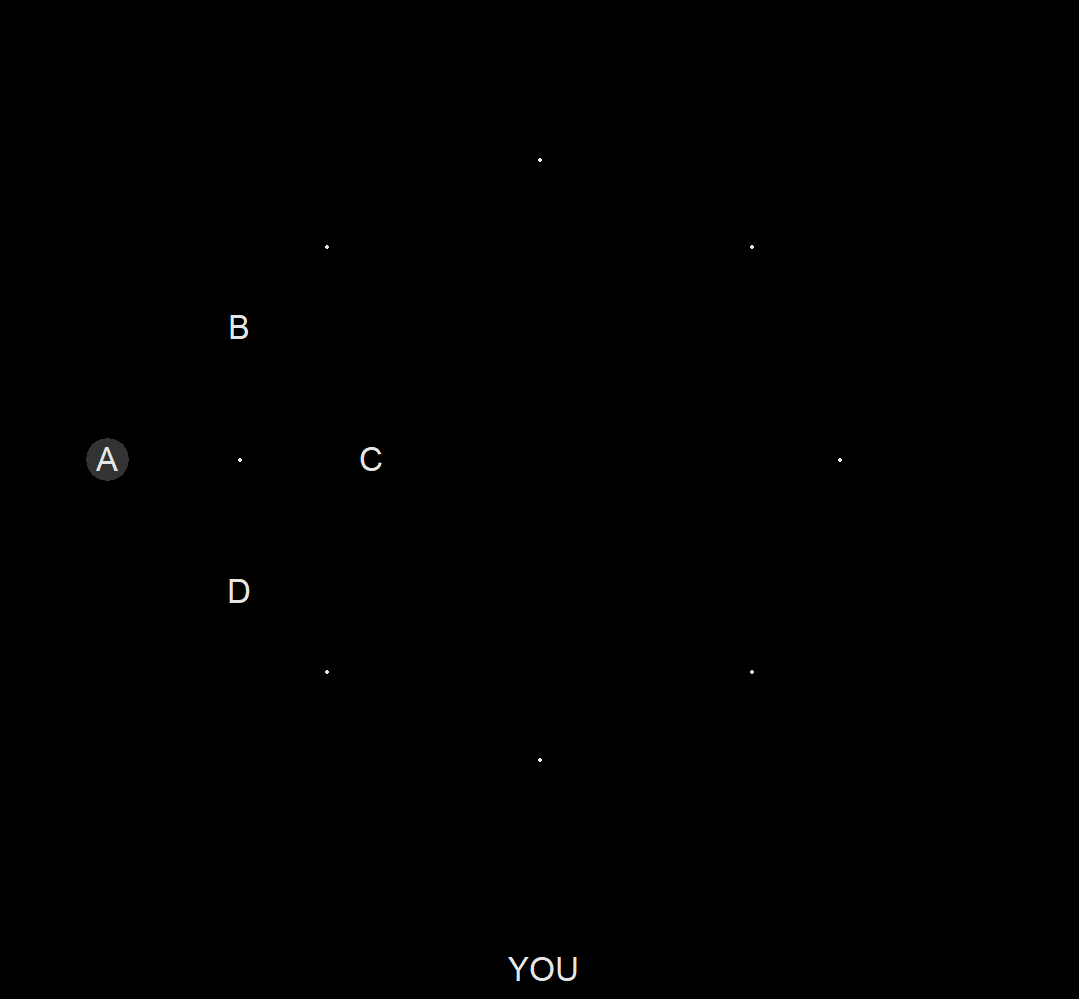}
    \caption{The hybrid eye typing interface. When the eyes look at the ABCD cluster, this cluster is highlighted as feedback (left). The characters in the selected cluster moved outwards (right).}
    \label{Fig.lookmovepre}
\end{figure*}

In order to overcome the aforementioned challenges, \cite{zeng_hybrid} proposed a hybrid eye typing interface (see Fig. \ref{Fig.lookmovepre}), which simplifies the calibration process and works with both low data accuracy and precision.  In this interface, the characters in the cluster being gazed at for a short period of time will move. To activate a command action, the user needs to follow the movement of the desired character in this cluster. This enables the recognition of activation based on the relative movement of the gaze position, rather than the absolute position. 
It can be configured in public settings as an alternative input method for special populations, such as those with physical disabilities, temporary physical limitations, individuals holding babies, or those with injured hands.
Besides, because cone density decreases with retinal eccentricity, visual acuity is generally lower in peripheral vision compared to foveal vision \citep{curcio1990human,pointer1989contrast}. This hybrid eye typing interface combines visual characteristics by distributing characters in positions that are relatively closer to the center of the field of view \ie foveal vision. Unlike hybrid eye typing interface, QWERTY has a rectangular layout and the characters on the sides require more eye movement to locate and select. Last but not least, In the hybrid eye typing interface, the gaze path required for activation is consistent when inputting each character.

While it overcomes some challenges, the maximum typing speed of the hybrid eye typing interface is limited due to its two-stage selection design. Hence, this study explores the use of predictive typing support to increase typing speeds for hybrid eye-typing interfaces.
The study has two main contributions:
\begin{enumerate}
    \item Investigation of the impact of letter and word prediction on typing performance and workload in hybrid eye-typing interfaces.
    \item Analyses of training time requirements for this hybrid eye-typing interface with letter and word prediction.
\end{enumerate}

\section{Related work}

\subsection{Dwell-based QWERTY eye typing layout}
In eye-typing, dwell-based QWERTY layout interfaces are the most frequently used interface design. In these, the user looks at the desired character for a predefined time to select it. 
To prevent the so-called \textit{Midas Touch} problem, where characters are selected, even though the user is just reading a letter, the dwell time is usually set longer than the duration of natural fixations, which range from 400-1000 ms \citep{majaranta2002twenty}. 
As an advantage, the learning cost for the QWERTY layout typing system is very low, as almost all users are familiar with this layout. 
As a major disadvantage, most eye-typing interfaces based on the QWERTY layout require calibration prior to use, as the character keys are grouped very closely together. 
As a related disadvantage, the accuracy of the gaze input registration can decrease during use, and re-calibration is needed to avoid unintentional selections \citep{feitchi2017}. 
To overcome the shortcomings, \cite{penkar2012designing} suggested that the size of each item of the eye typing interface should be designed to be relatively large.

\subsection{Alternative eye typing systems}
In addition to the dwell-based QWERTY eye typing interfaces, researchers have proposed alternative designs for gaze-only text entry interfaces.
For example, gaze gesture-based text entry interfaces, in which users draw shapes similar to letters, have been explored \citep{Wobbrock}. 
As a major alternative, multiple studies have researched dynamic eye typing interfaces, in which the size or position of characters may change throughout the interaction process, such as \textit{Dasher} \citep{ward2002fast}, and StarGazer \citep{10.1145/1344471.1344521}. 
Furthermore, researchers also focus on developing eye typing systems with a short calibration process or without calibration, such as gaze interfaces based on smooth pursuit eye movements \citep{lutz2015smoovs,10.1145/3314111.3319838,porta2021speye, EyeTell, zeng_hybrid}.  
In Table \ref{tab:paperlist}, we present a summary of studies focusing on two-stage eye-typing interactions that do not utilize the QWERTY keyboard layout.

\begin{table}[]
\centering
\caption{\label{tab:paperlist}Summary of the studies of non-QWERTY two-staged eye typing interfaces}

\begin{adjustbox}{width=1\textwidth}
\begin{threeparttable} 
\begin{tabular}{llcccll}
\toprule
Studies   & Predictive & Calibration & N & Phrases &WPM & Paper \\

&  function\\ \midrule
SMOOVS                                                                    & no   pred.                          & one-point                        & 24                                & same holoalphabetic German sentence for each condition& 2.90-3.34                &   \cite{lutz2015smoovs}    \\ 
SMOOVS                                                            & 1-word pred.                        & one-point                        & 6            &5 phrases                & 4.50                     &    \cite{zeng2018text}   \\ 
SMOOVS                                                               & 6-word pred.                       & free                             & 26                            & 5 phrases    & 3.41                     &   \cite{10.1145/3314111.3319838}    \\ 
EyeTell                                                               & letter pred.                       & free                             & 36                       & 4 trials (each contains 4 words)         & 1.27                     &     \cite{EyeTell}  \\
SPEye & no pred. & free & 32 &5 words& 0.85 & \cite{porta2021speye} \\
   & 5-word   pred.  & free     & 32       & 3 sentences& 1.15      &       \\
Hybrid interface       & no pred.                           & one-point                        & 29                       & 20 phrases        & 4.70                     &    \cite{zeng_hybrid}   \\ 
 \bottomrule
\end{tabular}
    \begin{tablenotes}                 
    \footnotesize               
    \item[1]five-day training
    \end{tablenotes}            
    \end{threeparttable}       
\end{adjustbox}
\end{table}

\subsection{Language model in eye typing}
Language models are used to improve typing performance by predicting the next word or letter, particularly in typing methods with slow input efficiency, such as brain-computer interfaces (BCI) \citep{orhan2012improved,speier2017online} and eye typing interfaces \citep{ZhangXuang,10.1145/2168556.2168605}, where text entry speeds are constrained by the input modality.
Nevertheless, the addition of a language model does not ensure a definitive enhancement in typing efficiency. Increasing prediction suggestion reduces keyboard actions but negatively impacts average time performance due to increased attention and usage costs \citep{10.1145/2858036.2858305}.  
Below we summarize studies related to how eye-typing interfaces can use language models to improve typing performance. 

\cite{ZhangXuang} introduced a weight-based algorithm for QWERTY layout eye typing, and compared letter and word prediction. 
With weighting factors, the system recognized the input letter based on the combination of word stem frequency and the distance from the current gaze position to adjacent letters. In an interface with letter prediction, the next three possible letters were given and highlighted, and the word prediction function was tested with five candidate words. 
They found that letter prediction proved to be an effective approach for enhancing typing performance in interfaces featuring small actionable objects. Word prediction exhibited only small superior typing performance in interfaces equipped with larger buttons.
Apart from \cite{ZhangXuang}, research utilizing language models to improve eye typing performance has focused on dwell-free eye typing methods \citep{10.1145/2168556.2168605}. 
In such a word-level typing system, users do not have to select letter by letter. 
Instead, the current typing word is predicted according to the gaze path that the user glances at the characters comprising the word. 
Since the eye is always on, this swipe-based method requires the user to specify the beginning and the ending of a word. \cite{10.1145/2858036.2858335} used eye gestures to distinguish them. Some studies tried to distinguish the beginning and the end by the bi-modal method, which combined touch and gaze input \citep{10.1145/3313831.3376317}.

In addition to the traditional QWERTY layout, there are also new eye typing interfaces containing word predictions, such as \textit{pEYEwrite} \citep{10.1145/1743666.1743738}, \textit{GazeTalk} \citep{hansen2003language}, \textit{SMOOVS} \citep{zeng2018text}, \textit{SliceType} \citep{benligiray2019slicetype}, and SPEye \citep{porta2021speye}. 
In the study SPEye, the typing performance with or without word prediction function was compared. \cite{porta2021speye} reported typing speed, measured in words per minute (WPM), increases from 0.85 WPM to 1.15 WPM when participants use word prediction. It also introduces more errors during typing, with the median typing errors (ERR, \ie errors incorrectly entered and not corrected) increasing from 0 to 0.01 and the median keystrokes per character KSPC increasing from 1 to 1.03. However, this study only provided a basic comparison between word prediction and the absence of word prediction.

Some studies have abandoned the keyboard interface, the most famous of which is \textit{Dasher} \citep{ward2002fast,ward2000dasher}. In \textit{Dasher}, the next possible letters move from the right to the left of the screen. Due to the typing system's strong dependence on the language model, users may encounter difficulties when attempting to input words that are not included in the existing corpus.

To sum up, despite advancements in eye typing interfaces, there remains a significant gap in the comprehensive exploration of integrating language models to enhance typing efficiency specifically for non-QWERTY layout eye typing interfaces. Thus, further research is warranted to delve into this area and uncover the potential benefits of leveraging language models in improving text input speed and accuracy in such interfaces.

\section{Hybrid eye-typing interface developed in this study}



This study proposes integrating letter and word predictions into the innovative layout of the two-stage eye typing interface to optimize typing performance. 
Additionally, we conducted a user test to assess typing performance and workload.

\subsection{Interaction design}

In the initial circle-layout eye typing interface (see Fig. \ref{Fig.lookmovepre}), users are able to select a character by initially directing their gaze toward the corresponding cluster, and subsequently tracking the movement of the desired character with their eyes. Additionally, the interface incorporates an idle area located at the center of the screen, where no actions are activated when the user's eye position resides within this region.



The text input interface is arranged in the order of 26 English letters, \ie A-Z, and the delete and space keys. The two function keys, \ie the delete (``DEL") and space keys (``SP") are in a group with the letters Y and Z. 
In particular, the interface is designed with eight clusters, each containing four characters. 
This design choice is driven by two main reasons:
1) When the user tracks one of the multiple objects moving in a straight line with his eyes, the target from interfaces containing both six and eight directions can be well recognized by the algorithm even though the accuracy of the eye tracking data is low \citep{zeng2020calibration}. Thus, an interface consisting of eight clusters, with each cluster moving in a distinct direction, was designed.
2) The amount of information one can obtain during a brief visual exposure is limited, \ie, within an exposure time of 15 to 500 milliseconds, the average number of letters correctly identifiable does not exceed 4.3~\citep{sperling1960information}.

The currently typed letters are displayed in the middle of the idle area. 
When the user reads the typed letters in the middle of the idle area, no action will be activated. 
If the user finishes typing a word, they can follow the ``SP" key, and the entered word will move from the center to the bottom of the screen. 

Before using the typing interface, there is a one-point calibration process. 
The process of entering one character can be divided into three phases:

\textbf{(0) Idle stage.} When the distance between the gaze point and the midpoint of the screen is less than the radius of the idle area, no action will be activated in this area. The detection happens only when the eye position is registered outside of the idle area. Whenever the eyes move back to the idle area during the interaction, all elements in the interface will gradually move back to their original positions, and the system returns to the initial idle stage until the user looks at the area beyond the inner idle circle.

\textbf{(1) Searching and cluster selection.} A cluster is detected when the angles of two consecutive gaze points fall within the recognition angle interval of this cluster. The selected cluster will be highlighted as immediate feedback, for example, the ``ABCD" cluster in Fig. \ref{Fig.lookmovepre}.

\textbf{(2) Character selection.} In a previous study \citep{zeng_hybrid}, participants reported that the moving speed (approximately 10$^\circ$/s) of individual characters was too fast, especially when users first started interacting with the system. Additionally, the search time threshold (400 ms) set for searching desired character was low. Hence, in this study, the search time threshold was set to 600 ms \ie if one cluster is looked at for longer than 600 ms, the characters in this cluster start to move outward. The default moving speed is about 250 pixels/s (equivalent to 6.4$^\circ$/s). To avoid distraction, when the selected cluster is moving outward, other clusters fade out. Fig. \ref{Fig.lookmovepre} presents the position of the characters when the movement of one cluster is finished.
The detection method relies on the measurement of angles. 
The function $\theta = \arctan2\left ( y_{g}-y_{m},x_{g}-x_{m} \right)$ is used to calculate the angle, which refers to the angle between the positive x-axis and the ray to gaze point $(x_{g}, y_{g})$ and the middle point $(x_{m},y_{m})$.
A short beep and a gray circle (see letter H in Fig. \ref{Fig.letterpre} ) are provided as auditory and visual feedback after a character is selected.

\subsection{Letter and word prediction}

Based on the original hybrid eye typing interface~\citep{zeng_hybrid}, two new interfaces are designed, which use language models for typing prediction.
One integrates letter prediction, while the other integrates both letter and word prediction. 
Including the original interface, three interfaces are compared: (1) No prediction (NoP), which served as a baseline, (2) Letter prediction (LP), and (3) Letter + word prediction (L+WP).

\begin{figure*}
\centering 
\includegraphics[width=0.8\textwidth]{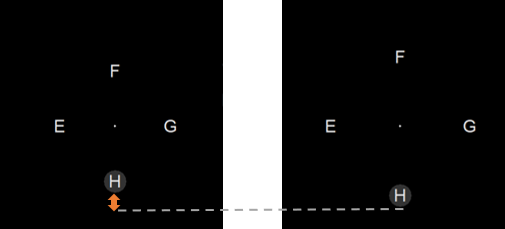}
\caption{The default value of the moving distance in phase (2) is 94 pixels (right figure), The figure on the left shows the moving distance when letter prediction is activated, \ie 68 pixels. }
\label{Fig.letterpre}
\end{figure*}

The interface with LP is visually similar to the NoP interface, as both contain seven clusters of actionable objects. 
The difference between NoP and LP is that the moving distance of some letters is decreased in specific cases, facilitating an easier selection. 
In the LP interface, if the letter belongs to the list of the next four most likely letters to appear, and the cluster containing this letter contains only one letter from the list, the moving distance for that specific letter is reduced by one-third in this cluster (see Fig. \ref{Fig.letterpre}).

\begin{figure*}
\centering 
\includegraphics[width=0.99\textwidth]{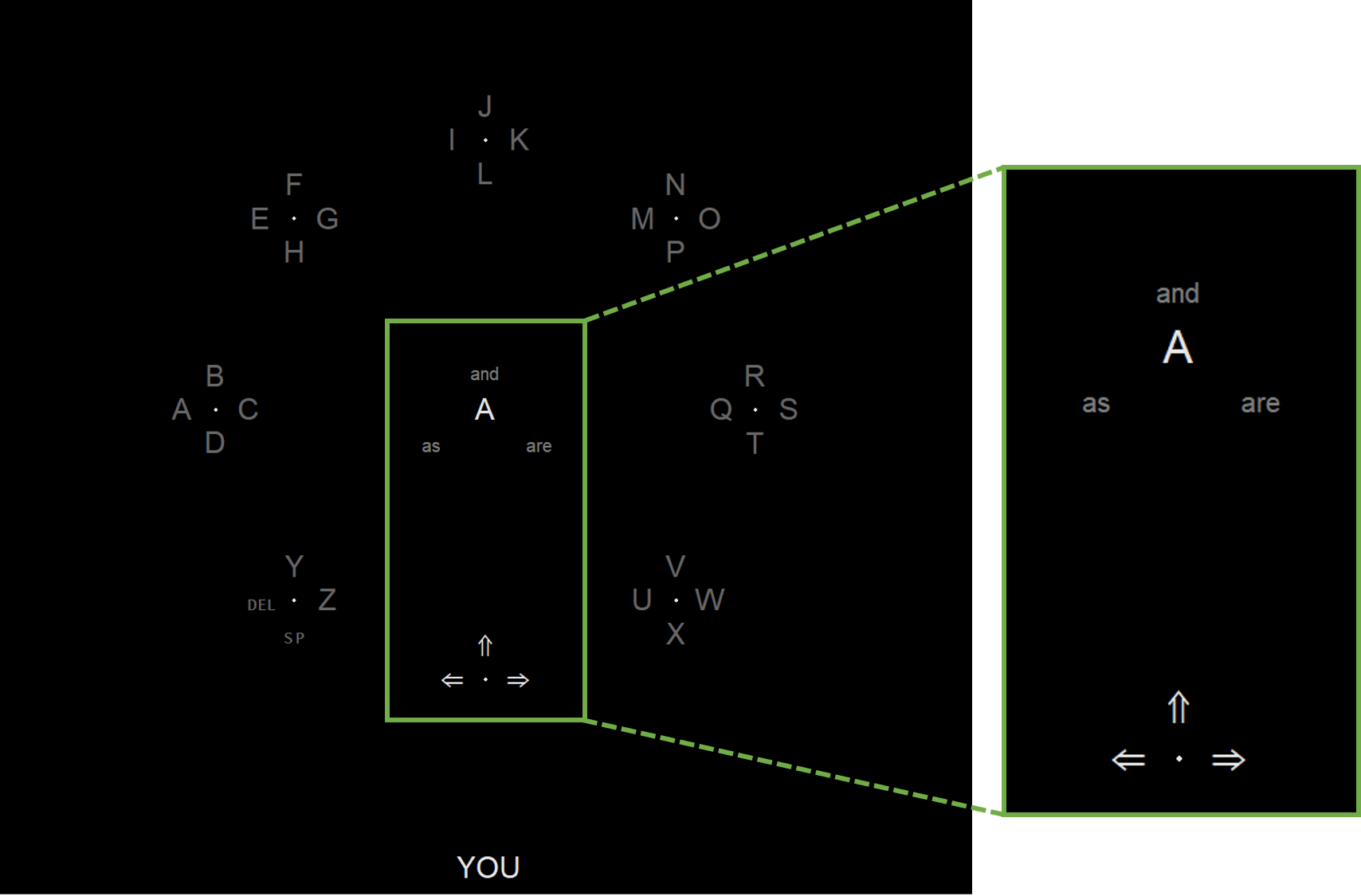}
\caption{The interface for L+WP. The user could follow the corresponding arrow to enter a predicted word. In this example, the up arrow stands for ``and", the left arrow represents ``as", and the right arrow is for ``are".} 
\label{fig:pre}
\end{figure*} 

The interface with L+WP contains eight clusters, which is one more cluster than the NoP and LP interfaces. In addition to the letter prediction function, two different word prediction functions are added, one predicting the currently typed word, and one predicting the potentially next typed word.
For the word prediction in the L+WP interface, a character-to-word model based on convolutional neural networks proposed by \cite{Kyubyong} is implemented. The network provides predictions for the (potentially) currently entered word, as well as the (potential) next word. 
Generally, predicted words are located in the central idle area, and when looking at these predicted words, no action will be triggered. In the cluster containing three arrows, each arrow corresponds to a predicted word in the corresponding position (see Fig. \ref{fig:pre}).
Since the entered words will be displayed at the bottom of the screen, no character is placed in the downward direction of the arrow cluster to prevent the user from triggering an unintentional selection when reading the entered words.
In L+WP, when a user is in the process of typing a word, three suggestions for word completion are given. In addition, when a word has been finalized, three candidates for the next possible words are provided.
15 more gaze points are collected to calculate the final angle for the arrow cluster. The arrows move a longer distance than other characters on the interface, resulting in an increased number of fixation points required for selection. Given the frequent usage of this cluster, particularly when word prediction is accurate, we endeavored to enhance the robustness of the word prediction function.

\section{Experiment 1}


\subsection{Experiment design}
The first experiment was constructed in a 3 (\textit{Prediction interface}) $\times$ 3 (\textit{Session}) within-subjects design. The factor \textit{Prediction interface} contained three levels (NoP, LP, L+WP) and three subsequent experimental \textit{Sessions} were implemented. Two phrases are typed in each session. The phrases were chosen from the phrases set proposed by \cite{10.1145/765891.765971}. All selected phrases were easy to understand and remember, such as ``you must make an appointment". The typed phrases vary from 16 to 32 with a mean of 24.67. The order of the three prediction interfaces was counterbalanced across participants. 
In total, 216 trials were run: 12 participants $\times$ 3 interfaces $\times$ 3 sessions $\times$ 2 phrases = 216 trials. 

\subsection{Evaluation metrics}
Words per minute (WPM), adjusted Words per Minute (AdjWPM), uncorrected error rate (UER), corrected error rate (CER), and keystroke savings (KS) were used to measure the typing performance. 
WPM was used to quantify the typing speed during text entry. One word is considered five characters, including letters, spaces, and punctuations. 
UER refers to how many errors remain in transcribed text. 
To penalize WPM based on uncorrected errors, we calculate the Adjusted Words per Minute ($AdjWPM= WPM * (1 - U)^a$), using a penalty exponent of 1.0 ($a = 1$), as participants can increase speed by allowing more errors.
CER is calculated by adding up the number of backspace key entries during text entry \citep{mackenzie2010text}. KS describes how many keystrokes are saved through the use of word prediction/completion \citep{trnka2008evaluating}.

The perceived workload was assessed using the NASA Task Load Index (NASA TLX). Besides, subjective feedback about the moving speed, search time, word prediction, and ideas for improvement was also collected through an open questions survey. 


\subsection{Participants}
Twelve participants (6 males, and 6 females) were recruited and the average age of the participants was 28.83 years ($SD$ = 4.88). Seven participants wore glasses, one wore contact lenses, and four did not wear any visual aids. 
Two participants knew about gaze-based interaction before, the others had no experience with gaze-based interaction. All participants were fluent English speakers, although none of them were native English speakers. Participants signed an \textit{Informed consent form} before starting the experiment and received monetary compensation after the experiment.

\subsection{Apparatus}
An eye tracker -- Tobii EyeX with 60 Hz was attached under a 24-inch monitor with a resolution of $1920 \times 1200$ pixels. The eye tracker was calibrated beforehand by the experimenter before the experiment. 
Participants were asked to sit in front of the monitor. 
If the participant's eye position moved out of the detection range of the eye tracker (too close or too far), participants were asked to adjust their position forward or backward until a detectable distance was reached. No other instructions were given. The distance between the user and the monitor ranged from 44 cm to 72 cm. At a distance of 60 cm between the participant and the screen, 1$^{\circ}$ was equivalent to 39 pixels.

\subsection{Procedure}

In experiment 1, the participants first signed an \textit{Informed consent form} and completed a short demographic questionnaire. Before the main experiment began, there was a brief introduction to the eye-tracking typing system, followed by a practice session. During the introduction, they received instructions on the layout and operation of the eye typing system. Subsequently, during the practice session, participants were provided with the opportunity to type with both typing interfaces, \ie with and without word prediction, allowing them to gain a thorough understanding of its functionality. They can practice typing some given words, such as “welcome”, and “hello” and short phrases like “Do you want coffee or tea?
At the beginning of each trial, there was a short one-point calibration. Then, the first phrase to be typed was displayed in the center of the screen. Participants were asked to remember this phrase and press the space key to start the typing task. Participants were asked to complete the NASA-TLX questionnaire after each session. Participants were instructed to enter the given phase as quickly and accurately as possible. They need to correct letters they typed unintentionally, not errors in predicted vocabulary. Participants were told that they could rest between trials. After the first phrase was typed, the second phrase was displayed and the same procedure was followed. After the second phrase, the experimental condition changed, and a new prediction condition was launched. 
The demographic questionnaire and open questions were answered after the text entry tasks. The whole experiment lasted approximately 40-60 minutes.

\subsection{Results}
\begin{figure}[tb] 
\centering 
\includegraphics[width=1\textwidth]{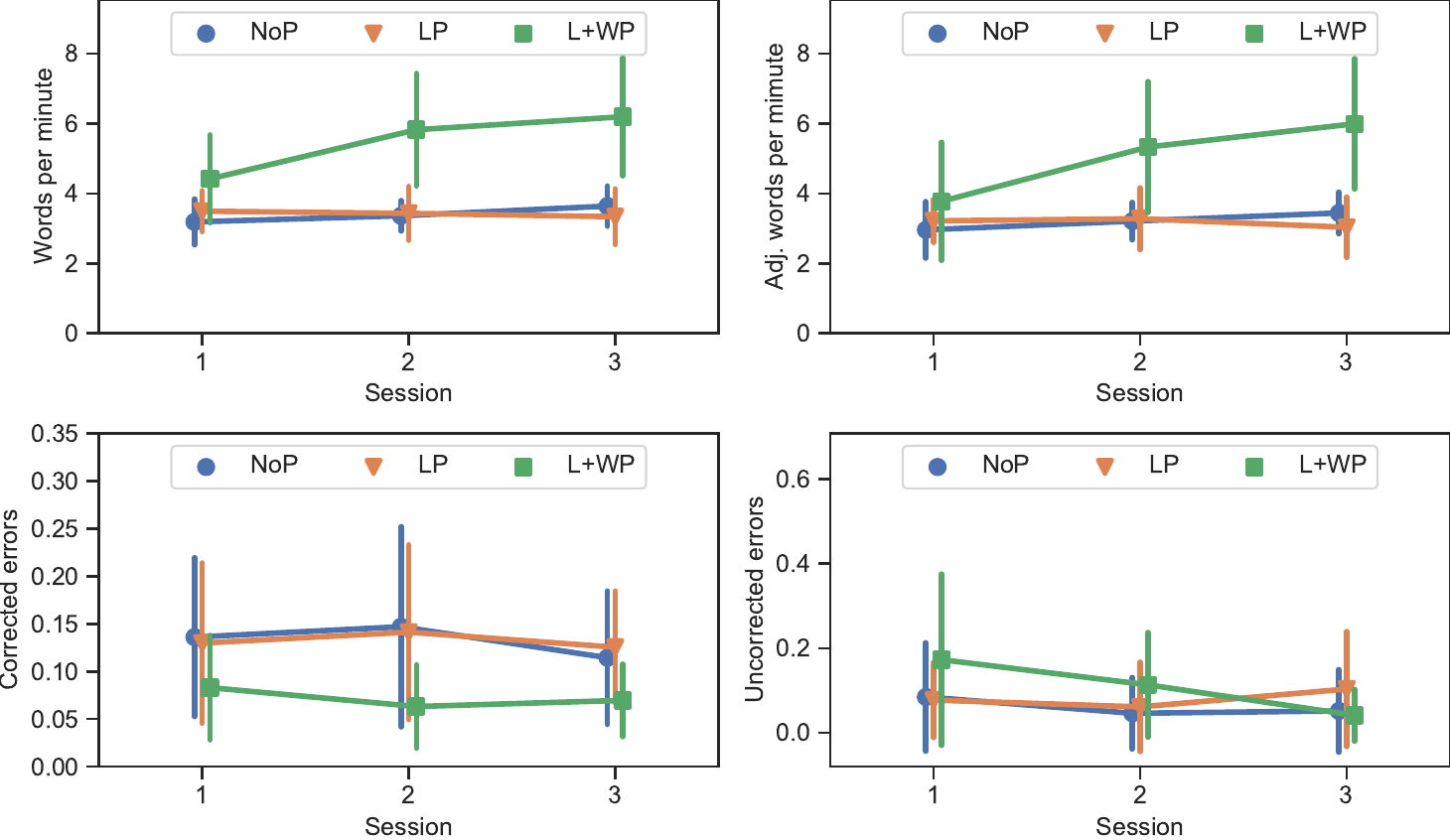}
\caption{Mean of WPM, AdjWPM, CER, and UER over 3 sessions (each session 2 phrases), the blue circle represents NoP, the orange triangle stands for LP, and the green square indicates L+WP (Error bars = $\pm$ 1~\textit{SD}).} 
\label{fig:objective_stu3}
\end{figure}

Repeated measures ANOVAs were used for data analysis, and pairwise comparisons were Bonferroni corrected. 
Shapiro-Wilk was used to check normality, and for non-normal data, the aligned rank transform (ART) was performed before ANOVA \citep{wobbrock2011aligned}.
Generalized eta squared $(\mathrm{\eta}_{G}^{2})$ was calculated as effect size \citep{10.3389/fpsyg.2013.00863}. Partial eta-squared $(\mathrm{\eta}_{p}^{2})$ was used for aligned ranks transformation ANOVA.
The descriptive statistics with regard to the objective evaluation are shown in Fig. \ref{fig:objective_stu3} and Table \ref{tab:tabst3}, and the ANOVA results are summarized in Table \ref{tab:st3sum_perfor} and \ref{tab:st3sum}. 

\subsubsection{Words per minute (WPM)}


\begin{table}[tb]
\caption{\label{tab:tabst3} Mean values ($M$) and standard deviations ($SD$) of WPM/AdjWPM, CER, and UER for each experimental condition.}
    \centering
    \begin{tabular}{ccccccccccccc}
    \toprule
        Methods & Sessions & \multicolumn{2}{c}{WPM} & ~ & \multicolumn{2}{c}{AdjWPM} & ~ & \multicolumn{2}{c}{CER} & ~ & \multicolumn{2}{c}{UER} \\ 
        ~ & ~ & $M$ & $SD$ & ~ & $M$ & $SD$ & ~ & $M$ & $SD$ & ~ & $M$ & $SD$ \\ \midrule
        L+WP & 1 & 4.42 & 1.31 & ~ & 3.77 & 1.76 & ~ & 0.08 & 0.06 & ~ & 0.08 & 0.06 \\ 
        ~ & 2 & 5.82 & 1.68 & ~ & 5.33 & 1.96 & ~ & 0.06 & 0.05 & ~ & 0.06 & 0.05 \\ 
        ~ & 3 & 6.19 & 1.77 & ~ & 5.99 & 1.95 & ~ & 0.07 & 0.04 & ~ & 0.07 & 0.04 \\ \addlinespace
        LP & 1 & 3.49 & 0.61 & ~ & 3.21 & 0.64 & ~ & 0.13 & 0.09 & ~ & 0.13 & 0.09 \\ 
        ~ & 2 & 3.43 & 0.8 & ~ & 3.28 & 0.92 & ~ & 0.14 & 0.1 & ~ & 0.14 & 0.1 \\ 
        ~ & 3 & 3.33 & 0.83 & ~ & 3.03 & 0.9 & ~ & 0.13 & 0.06 & ~ & 0.13 & 0.06 \\ \addlinespace
        NoP & 1 & 3.19 & 0.68 & ~ & 2.96 & 0.85 & ~ & 0.14 & 0.09 & ~ & 0.14 & 0.09 \\ 
        ~ & 2 & 3.36 & 0.45 & ~ & 3.21 & 0.56 & ~ & 0.15 & 0.11 & ~ & 0.15 & 0.11 \\ 
        ~ & 3 & 3.64 & 0.6 & ~ & 3.44 & 0.62 & ~ & 0.11 & 0.07 & ~ & 0.11 & 0.07 \\ \bottomrule
    \end{tabular}
\end{table}

A Shapiro-Wilk normality test showed that the WPM was normally distributed ($p>.05$).
Since Mauchly’s test indicated a violation of sphericity ($p<.05$), the Greenhouse Geisser correction was used to correct for the violation of the assumption of sphericity.
The results show a significant main effect of \textit{Prediction interface} on WPM $(F(1.19, 13.12) = 52.10, p<.001,$$\mathrm{\eta}_{G}^{2}=0.473$). 
A significant effect of \textit{Session} on WPM was also found $(F(2, 22) = 7.34, p<.01,\mathrm{\eta}_{G}^{2}=0.074)$.
There was a significant interaction between the \textit{Prediction interface} and \textit{Session} on WPM $(F(2.2, 24.2) = 4.95, p<.05, \mathrm{\eta}_{G}^{2}=0.104)$. 
The pairwise comparisons showed that there was a significant difference between sessions 1 and 2 ($p<.05$) and between sessions 1 and 3 ($p<.01$) regarding the L+WP interface. We see that the green line (L+WP) has a steeper increase from session 1 to session 3 whereas the orange (LP) and blue (NoP) lines are much more horizontal.
In addition, WPM in L+WP is significantly higher than in NoP across all three sessions ($p < .01$), and higher than LP during sessions 2\&3 ($p < .01$). 



\subsubsection{Adjusted words per minute (AdjWPM)}
The AdjWPM is normally distributed ($p>.05$) and sphericity assumptions were violated ($p<.05$). After Greenhouse Geisser correction, significant main effects were found in terms of \textit{Prediction interface} $(F(1.16, 12.74) = 33.43, p<.001,$$\mathrm{\eta}_{G}^{2}=0.343$) and \textit{Session} $(F(2, 22) = 7.14, p=.004,$$\mathrm{\eta}_{G}^{2}=0.080$). There was a significant interaction between \textit{Prediction interface} and \textit{Session} $(F(2.29, 25.20) = 4.44, p=.019,$$\mathrm{\eta}_{G}^{2}=0.110$). The pairwise comparison showed that L+WP is significantly higher than NoP in session 1 ($p < .01$), and L+WP is significantly higher than NoP and LP in sessions 2  ($p<.01$) and 3 ($p < .001$).

\subsubsection{Error rates}
\textbf{Corrected error rate (CER)} indicates how frequently the user corrects unintended or erroneously selected characters during typing.
A Shapiro-Wilk normality test showed that the CER was normally distributed ($p>.05$). Mauchly's Test for Sphericity was not significant regarding CER ($p>.05$). 
A significant main effect of the \textit{Prediction interface} was found ($F (2, 22)$ $ = 7.17, p<.01,\mathrm{\eta}_{G}^{2}=0.131$). There was no significant effect of \textit{Session} on corrected error rate $F (2, 22) = 0.76, p = .480,\mathrm{\eta}_{G}^{2}=0.008$. The interaction between these terms was not significant ($F (4, 44) = 0.67, p = .617,\mathrm{\eta}_{G}^{2}=0.011$). 
From the posthoc test, we found that the CER for the L+WP is significantly lower than that of the LP ($p<.001$ ) and the NoP ($p<.001$ ). However, no difference was found between LP and NoP.

\textbf{Uncorrected error rate (UER)} reports how many errors are left in the typed sentence. 
A Shapiro-Wilk normality test showed that the UER was not normally distributed ($p<.05$), thus, a two-way ART RM-ANOVA was performed. 
The results show a significant main effect of \textit{Prediction interface} on UER ($F(2, 88) = 3.45, p<.05,\mathrm{\eta}_{p}^{2}=0.073$). 
A significant effect of \textit{Session} on UER was also found $(F(2,88) = 3.88, p<.05,\mathrm{\eta}_{p}^{2}=0.081)$.
There was no significant interaction between the \textit{Prediction interface} and \textit{Session} on UER $(F(4, 88) = 1.94, p = .11,\mathrm{\eta}_{p}^{2}=0.081)$. 
Based on the posthoc test results, the UER for the L+WP is significantly lower than that of the NoP ($p < .05$), besides, the UER in session 3 is significantly lower than in session 1 ($p < .05 $).


\subsubsection{Keystroke savings (KS)}
\textbf{Keystroke savings} measures the percentage of key savings with the language model compared to letter-by-letter text entry. We analyzed the data for the L+WP interface. 
L+WP started with a KS of 0.38 ($SD = 0.2$) in session 1 to 0.44 ($SD = 0.08$) in session 2 and 3. 
A Shapiro-Wilk normality test showed that the KS was normally distributed ($p>.05$).
Since Mauchly’s test indicated a violation of sphericity ($p <.05$), the Greenhouse–Geisser correction was used to correct the violation of the assumption of sphericity.
The one-way ANOVA indicated that \textit{Session} had no significant influence on KS $(F(1.27, 13.95) = 1.07, p =.337,\mathrm{\eta}_{G}^{2}=0.047)$. 


\begin{table}[t]
\caption{\label{tab:st3sum_perfor}Statistical significance of main and interaction effects regarding performance measures.}
\begin{tabular}{lllllll}
\toprule
Measurement                      & Effect                        & $F$ & $df_{1}$ & $df_{2}$  & $p$ & $\mathrm{\eta}^{2}$ \\ \midrule
\multirow{3}{*}{WPM}   & interface  & 52.10   &   1.19 & 13.12& $p<.001$***  & 0.473     \\
                                 & session   & 7.34 &     2.00 & 22.00 & .004** & 0.074             \\
                                 &interface * session & 4.95& 2.20& 24.20  & .014*& 0.104 \\\addlinespace

\multirow{3}{*}{AdjWPM }   & interface  & 33.43   &   1.16 & 12.74& $p<.001$***  & 0.343     \\
                                 & session   & 7.14 &     2.00 & 22.00 & .004** & 0.080             \\
                                 &interface * session & 4.44& 2.29& 25.20  & .019*& 0.110 \\\addlinespace

\multirow{3}{*}{CER} & interface & 7.17 & 2 & 22& .004**  & 0.131           \\
                                 & session & 0.76 & 2 & 22& .480    &   0.008     \\
                                 & interface * session & 0.67&   4 & 44 & .617  &     0.011    \\\addlinespace
\multirow{3}{*}{UER} &  interface           &  3.45 &  2  &   88 &.036*&    0.073   \\
                                 & session                        &   3.8&  2  &   88 &.024*&    0.081   \\
                                 & interface * session &   1.94&  4  &   88 &.110  &  0.081    \\\addlinespace
KS     & session &1.07 &    1.27 &13.95 & .337 &      0.047          \\
\bottomrule
\end{tabular}
\begin{tablenotes}    
    \footnotesize               
    \item[1]‘***’ .001, ‘**’ .01, ‘*’ .05
    \item[2] Keystroke savings only applied for data from letter and word prediction interfaces.
    \end{tablenotes}   

\end{table}

\subsubsection{Subjective Evaluation}

\begin{figure}[bt]
\centering 
\includegraphics[width=1\textwidth]{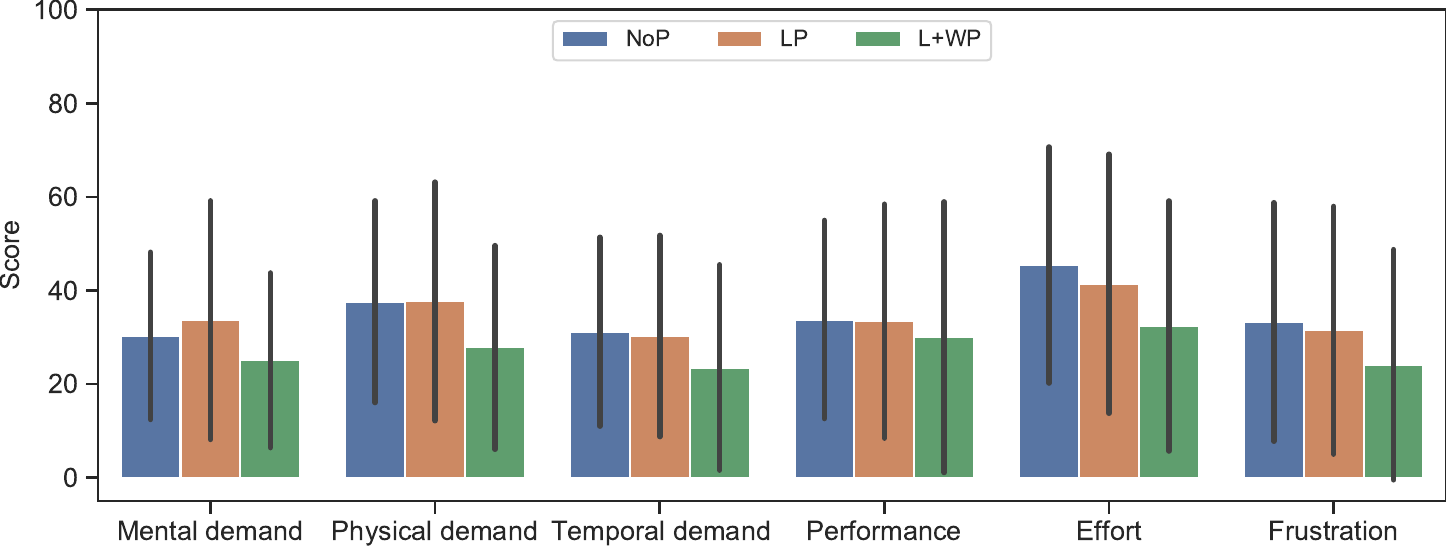}
\caption{Mean of NASA TLX results, lower scores represent lower workload (Error bars = $\pm$ 1~\textit{SD}).} 
\label{fig:nasa_stu3}
\end{figure}

After completing the task for each technique in each session, participants were asked to fill out the NASA TLX questionnaire. 
Fig. \ref{fig:nasa_stu3} illustrates the mean scores of the NASA Task Load Index in each dimension. 
Since these rating results were non-normal, the statistical significance tests were performed using a two-way ART RM-ANOVA.

The results reveal that descriptively, L+WP had a lower score than LP and NoP on all six dimensions.
There was a significant main effect of \textit{Prediction interface} on the mental, physical, temporal demand, and effort dimensions. 
A significant main effect was also found for \textit{Session} on performance and effort dimension. 
There was a significant interaction between the \textit{Prediction interface} and the \textit{Session} on mental demand. 

The results found that the mental demand decreased significantly between sessions~1 and~3 ($p<.001$) for L+WP, and decreased also significantly between L+WP and LP in session 3 ($p<.001$). Besides, significant differences were found between L+WP in session~3 and NoP in session 1 ($p<.001$), and between L+WP in session~3 and NoP in session 2 ($p<.05$).
In terms of physical demand, there were significant declines between L+WP and LP~($p<.05$ ), and between L+WP and NoP~($p<.05$).
For temporal demand, there was a significant decline between L+WP and NoP~($p<.05$).
The effort score for L+WP was significantly lower than NoP~($p<.05$).
Participants felt that they performed more successfully in session~3 than in session~1~$(p <.05)$.
There were significant decreases in effort between sessions~1 and~2 $(p<.05)$, and between sessions~1 and~3 $(p<.05)$. 





\begin{table}[t]
\caption{\label{tab:st3sum}Statistical significance of main and interaction effects regarding subjective measures.}
\begin{tabular}{lllllll}
\toprule
NASA-TLX                      & Effect                        & $F$ & $df_{1}$ & $df_{2}$  & $p$ & $\mathrm{\eta}^{2}$ \\ \midrule
\multirow{3}{*}{Mental demand}   & interface          & 3.46&  2 &    88& .036*      &  0.073   \\
                                 & session                        & 1.50 &  2  &   88 &.229          &  0.033   \\
                                 & interface * session &  3.18&  4 &    88 &.017*       &   0.126  \\\addlinespace
\multirow{3}{*}{Physical demand} & interface           &    6.52&  2   &  88& .002**        &  0.129   \\
                                 & session                        &    0.8&  2  &   88 &.432     &   0.019  \\
                                 & interface * session &     1.42&  4  &   88 &.232          &   0.061  \\\addlinespace
\multirow{3}{*}{Temporal demand} & interface           &    3.60&  2   &  88& .031*        &   0.076  \\
                                 & session                        &    0.32&  2  &   88& .725         &   0.007  \\
                                 & interface * session &     1.24 &  4 &    88& .300        &  0.053   \\\addlinespace
\multirow{3}{*}{Performance}     & interface           &     0.88&  2  &   88& .416         &  0.020   \\
                                 & session                        &    3.30&  2  &   88& .042*         &  0.070   \\
                                 & interface * session &    1.87& 4   &  88 &.122         &  0.079   \\\addlinespace
\multirow{3}{*}{Effort}          & interface           &     5.26&  2  &   88& .007**         & 0.107    \\
                                 & session                        &     4.29&  2  &   88& .017*       &  0.089   \\
                                 & interface * session &     1.77&  4  &   88 &.141        &  0.075   \\\addlinespace
\multirow{3}{*}{Frustration}     & interface           &      2.65&  2 &    88 &.076         &  0.057   \\
                                 & session                        &     0.15&  2  &   88 &.859         &  0.003   \\
                                 & interface * session &     1.35& 4  &   88 &.257          &  0.058  \\
\bottomrule
\end{tabular}
\begin{tablenotes}    
    \footnotesize               
    \item[1]‘***’ .001, ‘**’ .01, ‘*’ .05
    \end{tablenotes}   
\end{table}

\subsection{Qualitative feedback}
\noindent\textbf{Moving speed of characters}
More than half of the participants found the moving speed of characters appropriate. Four participants thought that the moving speed is too fast, and one of them reported that ``\textit{the moving speed is too fast only at the beginning of the experiment, and I can adapt to this speed after a period of practice.}" 
Only one participant assessed the speed as too slow.

\noindent\textbf{Searching time for the desired character}
Four participants felt that they had enough time to search for the desired letter. 
The other eight participants reported that they did not have enough time only at the beginning, but after practicing a few phrases, they could have time to find the desired letter.

\noindent\textbf{Views on word prediction}
Participants were asked three questions regarding their views on word prediction. The first question was about the intuitiveness of design. When the predicted word is visually distinguished from the activating arrow, can the user quickly understand and select the corresponding arrow. All participants considered that it was effortless to understand, and one of them suggested adding color for different predicted words. The second question was whether each arrow can be selected successfully. Six participants thought that they could select the arrows successfully, and one of them reported that arrows were selected earlier than letters. On the other hand, six participants reported that predictive words were easy to choose by mistake. The third question assessed whether the word prediction was helpful. Almost all users responded that vocabulary prediction is very helpful. One participant appreciated the word prediction, especially for lengthy words.

\noindent\textbf{Suggestions}
Five participants recommend separating the space and delete keys because they felt that misselection often occured between these two keys. 
Four participants suggested reducing the sensitivity of the system to reduce unintentional selections.

\section{Experiment 2}
In addition to the experiment 1, the potential effects of repeated practice with the system, as well as the variation of the moving speed of the interface were explored in this exploratory pilot user study. We iterated the interface based on the feedback from experiment 1. The character ``X'' and ``DEL'' \ie delete key swapped positions according to the subjective feedback of the users from experiment 1 (separate the delete and space key, see Fig. \ref{fig:pre} for comparison). 
\subsection{Experiment Design}
A two-factor mixed design was implemented, with \textit{Moving speed} of the interface as a between-subject factor, and the \textit{Session} as a within-subject factor. 
Two moving speeds were tested, a slower moving speed: 6.4$^\circ$/s (consistent with the speed condition of experiment 1) and a faster-moving speed:~10$^\circ$/s. To investigate potential practice effects, the number of sessions was increased to 10. 
Each individual participant entered a total of 14 phrases, \ie per participant, a total of 140 phrases were typed.  
A maximum of two sessions were tested per day with the minimum time between two sessions defined as two hours.
The dependent variables were words per minute (WPM), uncorrected error rate (UER), corrected error rate (CER), and NASA Task Load Index (NASA TLX). 

\subsection{Participants}
A total of 12 participants (9 male, 3 female) were recruited for the experiment 2, with ages ranging from 22 to 33 years old ($M$=27.9).
Six of them were selected for the fast-moving speed condition and six were tested with slow-moving speed. Hence in total, 12 participants $\times$ 10 sessions $\times$ 14 phrases = 1680 phrases were collected.
All participants were fluent English speakers. 
Four participants wore glasses, and one wore contact lenses. Only one participant had experience with eye-tracking devices. None of them had any experience with eye typing.

\subsection{Procedure}
The experimental setup was similar to experiment 1. After welcoming the participants, they were introduced to the eye typing interface, including the layout of the typing interface, how to select letters, and word prediction. There was no practice session before the first session. Participants were instructed to enter the given phase as quickly and accurately as possible. They were asked to rectify any unintentional letters but not errors within the predicted words.
Then, they were informed that there were 14 phrases for each session. 
The order of phrases was randomized. 
Each test session lasts about 20 - 30 mins.
After each session, participants were asked to fill out the NASA TLX questionnaire. 
After completing the last session, all participants received monetary compensation.

\subsection{Results}

\begin{figure}[htb]
\centering 
\includegraphics[width=1\textwidth]{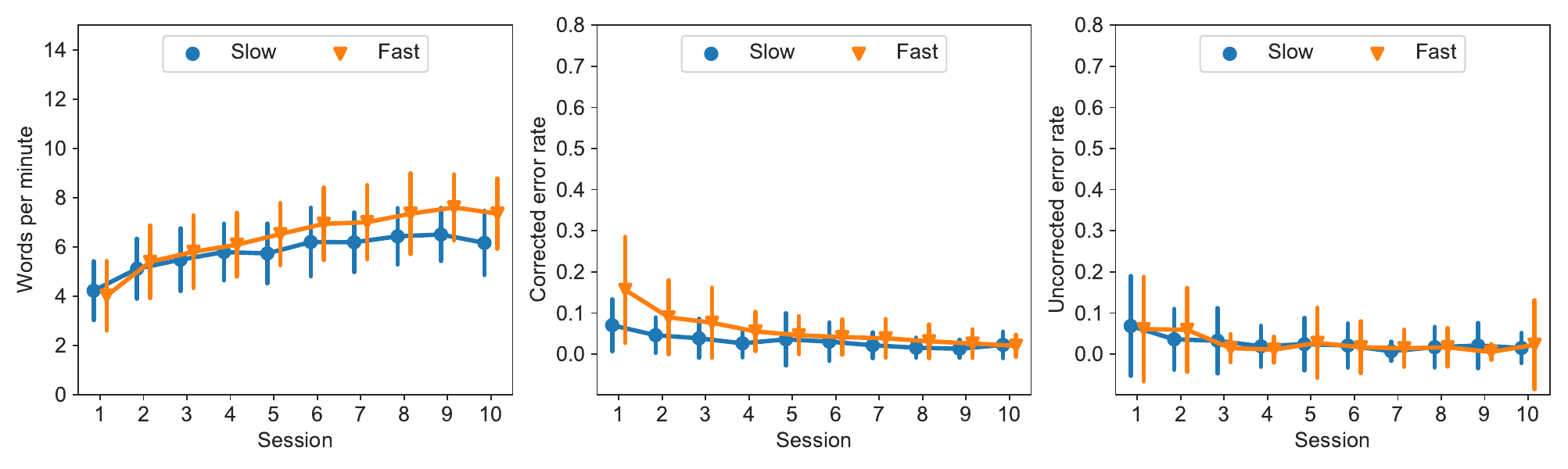}
\caption{Mean of WPM, CER, and UER over 10 sessions (each session 14 phrases), the blue circle represents slow-moving speed, and the orange triangle stands for fast-moving speed (Error bars = $\pm$ 1~\textit{SD}).} 
\label{fig:obj_stu_long}
\end{figure}

The descriptive results for WPM, CER, and UER are shown in Fig. \ref{fig:obj_stu_long}. 
Shapiro-Wilk normality tests showed that WPM, CER, and UER were normally distributed ($p> .05$). Levene’s test was used to evaluate the equality of variances.
The subjective rating results were non-normal, and the statistical significance tests were performed using a two-way ART RM-ANOVA.

\subsubsection{Words per minute (WPM)}
The average typing speed was 4.22 WPM in session 1 and increased to 6.17 in session 10 in the slow group, whereas the average typing speed was 4.02 WPM in session 1 and increased to 7.35 in session 10 in the fast group.
The results showed a significant main effect of \textit{Session} on WPM $(F(9,90) = 46.18, p <.001,\mathrm{\eta}_{G}^{2}=0.653)$. 
There was no significant effect of \textit{Moving speed} on WPM $(F(1,10) = 4.24, p = .067),\mathrm{\eta}_{G}^{2}=0.201$. 
A significant interaction was found between factor \textit{Session} and factor \textit{Moving speed} $(F(9,90) = 2.701, p <.01,\mathrm{\eta}_{G}^{2}=0.099)$.
The posthoc test showed that there was a significant difference between the fast and slow group for sessions 5 ($p <.05$), 8 ($p <.05$), 9 ($p <.01$), and 10 ($p <.01$).

\subsubsection{Error rates}

\textbf{Corrected error rate (CER)} \\
The average CER was 0.07 in session 1 and decreased to 0.02 in session 10 in the slow group, whereas the average CER was 0.16 in session 1 and decreased to 0.02 in session 10 in the fast group.
There were statistically significant main effects of \textit{Session}~$(F(1.79,17.85) = 11.73,p <.001,\mathrm{\eta}_{G}^{2}=0.463)$ and \textit{Moving speed}~$(F(1,10) = 7.45, p <.05,\mathrm{\eta}_{G}^{2}=0.166)$ on CER. No significant interaction effect was found~$(F(1.79,17.85) = 2.27, p = .137,\mathrm{\eta}_{G}^{2}=0.143)$.~The posthoc test showed a significant decrease between sessions 1 and 2 ($p<.01$).



\noindent\textbf{Uncorrected error rate (UER)} \\
The average UER was 0.07 in session 1 and decreased to 0.01 in session 10 in the slow group, whereas the average UER was 0.06 in session 1 and decreased to 0.02 in session 10 in the fast group.
The results showed a significant main effect of \textit{Session} on UER $(F(2.77,27.69) = 46.18, p <.01,\mathrm{\eta}_{G}^{2}=0.355)$.
No significant differences were found in terms of \textit{Moving speed} $(F(1,10) = 0.03, p=.862,\mathrm{\eta}_{G}^{2}=0.001)$. And the interaction is not significant $(F(2.77,27.69) = 0.94, p = .427,\mathrm{\eta}_{G}^{2}=0.065)$.
The posthoc test showed significant differences between sessions 2-3 ($p<.05$).


\subsubsection{Subjective Evaluation}
\begin{figure}[!ht]
\centering 
\includegraphics[width=1\textwidth]{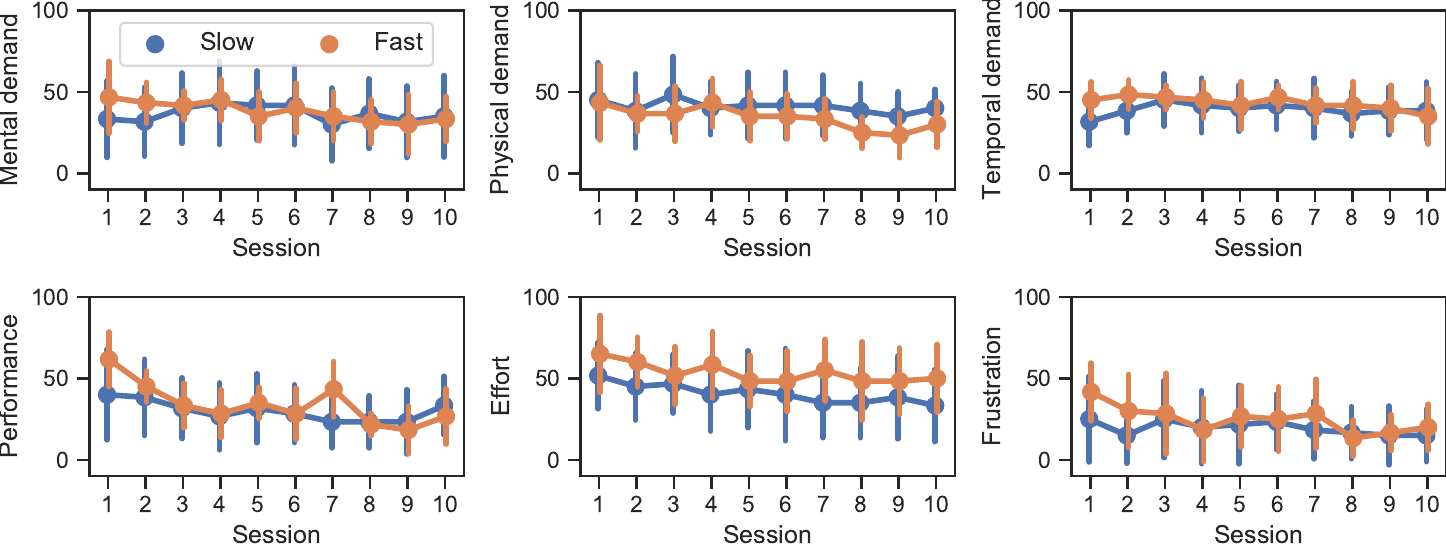}
\caption{Mean of NASA TLX results of experiment 2, lower scores represent lower workload (Error bars = $\pm$ 1~\textit{SD}).} 
\label{fig:nasa_e2}
\end{figure}

In terms of NASA TLX measurements, the average task load for each dimension is shown in Fig. \ref{fig:nasa_e2}. Except for the physical demand dimension, fast-moving speed received a higher rating.
There was a significant main effect of \textit{Session} on Physical demand $(p<.05)$, performance $(p<.001)$, effort $(p<.05)$ and frustration $(p<.05)$ dimensions. The Ratings for those dimensions reduced in fluctuations. 
No significant main effect was found for \textit{Moving speed} on NASA TLX scores. 
No significant interaction effect was found.

\section{Discussion}
\subsection{Findings from the experiment 1}
The experiment 1 compared three typing interfaces, and the results were analyzed in terms of text entry performance, NASA TLX, and qualitative feedback. 
Fig. \ref{fig:objective_stu3} shows that there was a significant improvement in typing speeds in both words per minute and adjusted words per minute (remove all uncorrected errors in the transcribed text due to unintentional activation of predicted words) by adding the word prediction. To facilitate comparisons with other studies, the discussion here primarily revolves around words per minute.
With the provision of the word prediction function, the words per minute were significantly faster than LP and NoP interfaces. 
The typing speed of L+WP increased greatly from session 1 to session 3, while the typing speed of LP and NoP increased more modestly. 
The corrected error rate of L+WP was significantly lower than LP and NoP. 
Using word prediction reduces the number of keystrokes required for users to enter phrases of a certain length, and at the same time, reduces the possibility of unintentional input to decrease the number of corrections correspondingly.
For the uncorrected error rate, there was a significant difference in prediction interfaces, mainly between L+WP and NoP. 
In this experiment, participants were told that they don't have to correct the errors for word prediction which was incorrectly entered because we want to know whether the word prediction can be selected correctly. 
From the results, we found that the UER is relatively high for L+WP in sessions 1 and 2 compared to LP and NoP. 
However, the UER of L+WP gradually decreased over sessions.
The gap with NoP was already very small in session 3, and even lower than LP.
For beginners, the rate of unintentional selection of word prediction is relatively high, thus, the UER is relatively high in the first two sessions, however, those errors were significantly reduced with short practice sessions.
Besides, although there was no significant difference in KS over session in L+WP, the trend indicates that descriptively, more keystrokes were saved from sessions 1 to 2. 
The results showed that participants were able to learn the novel interface after a short practice, and the interface with word prediction also exhibited a more efficient typing speed. 
However, false activation of the prediction function was more frequent in sessions 1 and 2. That is to say, the participants need more time to get familiar with the interface with the word prediction function to reach a relatively stable text entry speed. 

The results of the NASA TLX test (see Fig. \ref{fig:nasa_stu3}) reveal that L+WP had a lower score than LP and NoP on the mental, physical, time, and effort dimensions. These disparities notably intensified from sessions 1 to 3. 
The participants evaluated that the performance was getting better and the efforts devoted to accomplishing the task were decreased over sessions. 
These results also confirm the findings of the text entry performance.
When the user is not familiar with the text entry system, the participants' cognitive load is relatively high, and they tend to focus on the task of finding letters; thus, they do not have enough cognitive resources to perceive the update of the predicted vocabulary.
The cognitive load of finding each desired character decreased after participants became familiar with the interaction. They had more cognitive resources for perceiving word prediction and relied more on word prediction.


In this study, we discovered that not all eye-typing designs integrated into language models yield substantial improvements. 
Contrary to our expectations, the inclusion of LP did not result in superior typing performance, including factors such as increased typing speed, reduced error rates (see Fig. \ref{fig:objective_stu3}), and lower workload index (see Fig. \ref{fig:nasa_stu3}) when compared to NoP. 
Similar findings have also been reported in a previous study, \citet{ZhangXuang} introduced weight based on a language model to assist the eye typing, and no significant improvement was found compared with the baseline in terms of typing speed and error rate. 
In addition, we also found the mental demand scores of L+WP and NoP were slightly decreasing with practice, however, the score of LP did gradually increase. 
Even in session 3, there was a significant difference between L+WP and LP on mental demand.

We see two potential reasons for this. 
First, similar to the findings of several studies, the rhythm of each key activation is very necessary to be taken into account, as maintaining a regular typing rhythm leads to a smooth typing experience \citep{majaranta2006effects,10.1145/3025453.3025517}.In the LP interface, the moving distance of the letters predicted with high probability was reduced by one-third. 
This difference might have interrupted the typing rhythm of the participants. For subsequent improvements, a more moderate approach for reducing the selection trajectory for letters in letter prediction interfaces appears necessary. 
Second, in the case of LP, a letter with a high prediction probability will be ``easier" to be selected. 
If the letter is what should be entered, its activation cost can be reduced. 
However, when the predicted letters do not match the intended ones, there is an increased likelihood of errors. Specifically, when entering letters with a low occurrence probability, there is a higher chance of unintentionally activating letters with higher occurrence probabilities due to the predictive model being trained on the high-frequency letters in the training set.

\subsection{Findings from the experiment 2}
In experiment 2, we discovered that the L+WP interface demonstrates ongoing improvement in typing speed, suggesting that it is still in a progressive phase compared to the other two interfaces examined in experiment 1.
Thus, experiment 2 investigated further how fast users can learn to use the eye typing system to reach a plateau of typing speed for interfaces with fast and slow-moving speeds.
The typing speed (see Fig. \ref{fig:obj_stu_long}) increased in both the slow and fast interfaces during the first three sessions, which is due in large part to a significant reduction in error rates. 
A significant difference in the corrected error rate was found between sessions 1 and 2, which means less unintentional activation since users use the delete key less often. 
The significant decrease in the uncorrected error rate between sessions 2 and 3 implies that users are more correctly using the word prediction at this time point, as the setting in the experiment was that the activated word prediction could not be modified. 
In terms of the moving speed, in session 1, the slow group has a slightly faster typing speed than the fast group. However, the typing speed of the fast group surpassed that of the slow group in session 2.
During the initial sessions, the disparity between the fast and slow groups was not significantly pronounced, while a significant difference in typing speed emerged between the slow and fast groups in session 5.
Although the fast group had a faster typing speed after session 2 (see Fig. \ref{fig:obj_stu_long}) and the results of workload are not significant between moving speeds (see Fig. \ref{fig:nasa_e2}), the CER of the fast group was higher than that of the slow group, except for session 10.
When users are new to or starting to learn this typing interface, a slower speed is good for them to master this typing interface. After some typing experience (e.g., 10 phrases), a slower speed may not satisfy the need for typing efficiency, and it is appropriate to increase the moving speed.

In summary, we propose a design with corresponding directional arrows representing predicted words in different directions, which is suitable for the word prediction function in eye-tracking typing. 
This design successfully addresses the specific challenge of determining the optimal dwell time required for perceiving and comprehending the predicted words. Separating predictive words is feasible because perceiving and understanding predictive words through eye gaze both requires time and cognitive resources \citep{10.1145/2858036.2858305}. 
The user study shows that users can quickly comprehend and select these alternatives, and this particular design does not increase the typing load for users. In fact, due to word prediction, user effort is reduced compared to both no prediction and letter prediction. In line with prior research findings, our study replicated the notion that letter prediction does not yield superior performance compared to typing without prediction assistance. To dig deeper, we recruited participants to practice typing on this system for five days. We then examined how their typing speed and error rates changed during this training period, giving us a better understanding of how people adapt to this typing method.

\subsection{Limitations and future works}
While the study contributes valuable insights into this aspect of calibration-friendly eye typing interface, it is important to acknowledge several limitations that should be considered in the interpretation of the findings. A more detailed analysis needs to be taken into account in data analysis, e.g., How frequently the user utilizes the delete key to amend unintentionally entered predictive words or letters. Furthermore, we can analyze how the frequency of these occurrences changes with practice. Besides, the usability of the eye typing interface is not tested in a real scenario due to resource limitations.
In future work, more user studies could be conducted in real-life scenarios, such as railway stations, museums, hospitals, and so on. 
As a purely eye-controlled interface, this typing interface can also be used in headset devices to input text, as it has a low learning cost and the distance to move for each character is basically isometric. 
For the interaction that exclusively utilizes cursor points for text input, the user experience could be better than that of the traditional QWERTY keyboard, which was originally designed for ten-finger typing.
Another possibility for future research is the exploration of combining gaze input with other input modalities to obtain more efficient and natural user experiences. 
For example, by using head movements~\citep{10.1145/3379155.3391312,10.1145/3448017.3457379}.

\section{Conclusions}
This paper introduced a contactless one-point calibration eye typing interface that can be used for scenarios where text input is performed in a public space. 
Three eye typing interfaces (L+WP, LP, NoP) were compared on typing performance and perceived workload. 
The results show that the provision of word prediction significantly outperformed the other two prediction interfaces (LP, NoP) in terms of typing speed and error rates. Besides, L+WP also got a better rating in terms of NASA TLX on subjective perceptions such as mental workload and effort. Additionally, we found, that higher interface speeds lead to increased error rates, which only disappear after a number of repeated interactions with the system.


\section*{Declaration of competing interest}
The authors declare that they have no known competing financial interests or personal relationships that could have appeared to influence the work reported in this paper. \\

\section*{About the Authors}
\textbf{Zhe Zeng} received her M.S. and Ph.D. degrees from the Technical University of Berlin, Germany. Her research interests include eye tracking, natural gaze interaction, and human-robot interaction.\\

\noindent \textbf{Xiao Wang} received his M.S. degree from the Technical University of Berlin. His interests include human-machine interface.\\

\noindent \textbf{Felix Wilhelm Siebert} received his PhD degree in psychology from the
Leuphana University of Lüeneburg, Germany. He holds an assistant
professor position for transport psychology at the Department of
Technology, Management and Economics at the Technical University
of Denmark. \\

\noindent \textbf{Hailong~LIU} is an Assistant Professor at Graduate School of Science and Technology, NAIST, Japan.
He received his Ph.D. degree in Engineering from Ritsumeikan University, Japan in 2018.
His research interests include HMI designs, driving behavior modeling, motion sickness modeling, and trust calibration for human-autonomous vehicle interactions.

\bibliographystyle{apacite}
\bibliography{interactapasample}

\end{document}